\documentclass[a4paper,12pt]{article}
\usepackage{graphicx}
\usepackage[greek,english] {babel}
\usepackage[utf8x] {inputenc}
\usepackage[version=3] {mhchem}
\usepackage {amsfonts}
\usepackage{array}
\usepackage{amsmath}
\usepackage[dvipsnames]{xcolor}
\usepackage{young}
\usepackage[export]{adjustbox}
\usepackage{afterpage}
\usepackage{easybmat}
\usepackage{physics}
\usepackage{authblk}
\usepackage{hhline}
\usepackage{geometry}
\usepackage{times}
\usepackage{placeins}
\usepackage{cite}

\begin{document}
\selectlanguage{english}


\title{\bf Magic numbers of cylindrical symmetry}

\author{Andriana Martinou}
\author{Dennis Bonatsos}
\affil{\footnotesize Institute of Nuclear and Particle Physics, National Centre for Scientific Research  ``Demokritos'', GR-15310 Aghia Paraskevi, Attiki, Greece}

\maketitle
\begin{center}
\rule{16.5cm}{0.3mm}
\begin{abstract}
\small
In nuclear physics a magic number is defined as the nucleon number, which is separated by a significantly large single-particle energy gap from the next nucleon. Magic numbers define the nuclear shells, which are considered to be active, only if they are partially occupied by nucleons. As a consequence the single particle interactions of the valence nucleons lead to the description of the collective properties of the whole nucleus in the shell model theory. But phenomena as the island of inversion, the shape coexistence and the break down of the $N=20$ magic number reveal that the above definition of a magic number is deficient. A complementary definition should rely on the selection rules of the single particle interactions. Specifically the selection rules of the quadrupole-quadrupole interaction lead to two sets of magic numbers, namely the harmonic oscillator magic numbers 2, 8 20, 40, 70, 112, ... and the 
spin-orbit SO-like magic numbers 2, 6, 14, 28, 50, 82, 126, ...  The underlying symmetries are respectively the spherical symmetry of the 3D isotropic harmonic oscillator and the cylindrical symmetry of the 3D anisotropic harmonic oscillator with two frequencies equal. The above two sets of magic numbers along with the Elliott SU(3) symmetry framework predict long standing and puzzling phenomena in nuclear physics. 
\end{abstract}

\rule{16.5cm}{0.3mm}
\end{center}

\section{Introduction}

Magic numbers were one of the first discoveries in nuclear structure \cite{Elsasser}. It was realized that nuclei formed by particular numbers of protons (Z) and neutrons (N) were extremely stable. This experimental discovery led to the concept of a nuclear closed shell \cite{Mayer1}, in analogy to atomic closed shells. As a result, the shell model was introduced as the basic microscopic model for the description of atomic nuclei, its magic numbers for either protons or neutrons being 2, 8, 20, 28, 50, 82 \cite{Mayer2,Jensen,MJ,Heyde,Talmi}. The next magic number for neutrons is known to be 126 \cite{Sorlin}, while the one for protons remains elusive and is currently the subject of intensive research in superheavy elements \cite{Oganessian1,Oganessian2,Agbemava,Prassa}.  

At the infancy of nuclear structure it has also been realized that excitation energies of the atomic nucleus can be arranged into bands \cite{Scharff} with increasing angular momentum $J$. Soon thereafter it was realized, that energy bands near closed shells, would increase almost linearly with angular momentum, i.e., $E(J)=AJ$, therefore called vibrational or spherical, while energy bands near the middle of the nuclear valence shells would increase almost following the rigid rotator expression $E(J)=AJ(J+1)$, therefore called rotational or deformed \cite{Casten}. From the macroscopic point of view these bands have been described in the framework of the Bohr Hamiltonian \cite{Bohr}, soon extended into the Bohr Mottelson model \cite{BM1,BM2}, in terms of the collective deformation parameters $\beta$ and $\gamma$, of which the former describes the degree of deviation of the nuclear shape from sphericity, while the latter describes its departure from spherical symmetry towards cylindrical symmetry realized through axial prolate shapes (elongated like the ball of american football) or axial oblate (pancake-like ) shapes, or even towards purely triaxial shapes \cite{Davydov,MtV}, in which the cylindrical symmetry is also broken. It should be noticed, that the Bohr Mottelson model \cite{Bohr,BM1,BM2,EG,Rohozinski} involves 5 dimensions, the two collective variables $\beta$ and $\gamma$ plus the three Euler angles describing the orientation of the atomic nucleus in space. In addition to the angular momentum $J$, nuclear bands are additionally characterized by the projection of the angular momentum on the z-axis of the coordinate system attached to the nucleus \cite{Casten,Maruhn}, which is called the intrinsic set of coordinates. In the case of even nuclei, the symbol $K$ is used for this quantum number. Bands with $K=0$ contain the angular momenta $J=0,2,4,6,8,\dots$. The lowest one is the band built on the ground state (which is always zero for even-even nuclei, i.e. nuclei with even number of protons and even number of neutrons), called the ground state band. For the rest $K=0$ bands, the term $\beta$ bands is used. In contrast, bands with $K=2$ contain the angular momenta $J=2,3,4,5,6,\dots,$ and are called $\gamma$ bands \cite{Casten,Maruhn}.

Symmetries, using the language and techniques of group theory \cite{Wybourne,Gilmore,Dawber1,Dawber2,Tung,Chen,IacLie}, have also been used for the description of atomic nuclei \cite{IA,IPVI,FPVI,book,Jolie,RW,Chacon3}. It has been realized \cite{WJ,Rakavy,Bes,Corrigan,Chacon1,Chacon2}, that the Bohr Hamiltonian possesses a U(5) symmetry, having an O(5) subalgebra, which in turn has an SO(3) subalgebra. The SO(3) subalgebra is needed within all symmetries describing atomic nuclei, since the quantum number determining its irreducible representations is the angular momentum $J$. Therefore, SO(3) has to be present, so that the labeling of energy levels by the angular momentum quantum number $J$ is possible. U(5) is counting the number of excitation quanta present, while the irreducible representations of O(5) are characterized by the seniority quantum number \cite{WJ,Rakavy,Bes,Corrigan} corresponding to the number of nucleon pairs coupled to nonzero angular momentum. The O(5) symmetry imposes certain degeneracies in the spectrum. For example, the first excited $J=2$ state has to be degenerate with the first $J=4$ state, the first excited $J=4$ and $J=3$ states have to be degenerate to the first $J=6$ state, and so on (see Table I and Fig. 3 of Ref. \cite{U5E5} for further details). Since the seniority quantum number appears to be meaningful near closed shells \cite{Talmi}, O(5) is a symmetry expected to appear there.

For deformed nuclei, the SU(3) symmetry, possessing an SO(3) subalgebra, has been used. In addition to the three components of the angular momentum, forming the SO(3) subalgebra, it also contains the five components of the quadrupole operator. Its relevance to nuclear spectra has been demonstrated by Elliott \cite{Elliott1,Elliott2,Elliott3,Harvey} and will be further discussed in Section 4. 

An algebraic model incorporating U(5) and SU(3) as special subcases is the Interacting Boson Model \cite{IA,IPVI,FPVI,book,Jolie,Chacon3} possessing an overall U(6) symmetry. In addition to the U(5) \cite{AI1} and SU(3) \cite{AI2} subalgebras, it also contains an O(6) \cite{AI4} subalgebra, which in turn contains an O(5) and an SO(3) subalgebra. The O(6) symmetry \cite{AI4} turns out to be applicable to $\gamma$-unstable nuclei, i.e. to nuclei which can change easily their axial or triaxial character at the expense of no energy. Since the O(6) symmetry contains the same O(5) and SO(3) subalgebras as the U(5) symmetry, they exhibit the same set of degeneracies in their bands. In the Interacting Boson Model the nucleus is approximated by a set of bosons corresponding to correlated valence proton pairs and valence neutron pairs, i.e. nucleon pairs outside closed shells. But symmetries appear also in the fermionic description of the nucleus \cite{RW}, as we shall see below.

It should be noticed, that the study of the three-dimensional harmonic oscillator (3D-HO) in spherical coordinates leads to shell model algebras possessing SU(3) subalgebras \cite{Wybourne,MS,BK}. We shall use the spectroscopic notation, in which angular momenta $J=0,1,2,3,4,5,6,\dots$ correspond to the symbols s, p, d, f, g, h, i, \dots respectively. The p shell closes a U(3) algebra, the sd shell closes a U(6) algebra, the pf, sdg, pfh, and sdgi shells close U(10), U(15), U(21), and U(28) algebras, all of them possessing SU(3) and SO(3) subalgebras \cite{BK}. The s shell in the beginning of this series possesses a U(1) algebra. Since each value of the angular momentum can accommodate 2(2J+1) nucleons of the same type (protons or neutrons), the factor of two due to the two possible orientations of spin, one can easily work out the magic numbers appearing in the 3D-HO, which are 2, 8, 20, 40, 70, 112, 168, \dots.   

One can remark that the shell model magic numbers and the 3D-HO magic numbers are identical up to 20. This means that up to the sd shell, the SU(3) symmetry of the 3D-HO oscillator is present, as first realized by Elliott \cite{Elliott1,Elliott2,Elliott3,Harvey}. The agreement between the shell model magic numbers and the 3D-HO magic numbers is destroyed beyond the sd shell by the spin-orbit interaction, which is a relativistic effect which has to be introduced by hand in nonrelativistic shell model Hamiltonians in order to establish agreement with the experimentally observed magic numbers\cite{Mayer2,Jensen,MJ,Heyde,Talmi}. Various approximation methods have been developed over the years, attempting to reestablish the SU(3) symmetry beyond the sd shell. Pseudo-SU(3) \cite{pseudo1,pseudo2,DW1,DW2,Ginocchio} and quasi-SU(3) \cite{Zuker1,Zuker2} were the earliest ones, followed recently by proxy-SU(3) \cite{proxy1,proxy2,proxy3}, to be considered in more detail in Section 5. The search for approximate SU(3) symmetries beyond the sd shell has been stimulated by the tremendous amount of existing analytical and mathematical work in SU(3), which in this way would become available for use in the nuclear structure framework. 

Following a different path, deformed nuclear shapes have been described since a very early stage within an elementary shell model, using a three-dimensional anisotropic harmonic oscillator with cylindrical symmetry (occurring when two of the three frequencies are equal to each other), to which the spin-orbit interaction has been added. This is the Nilsson model \cite{Nilsson1,Nilsson2}, which over the years has been extremely useful in classifying and understanding a huge bulk of experimental data \cite{RN,Mottelson}. The Nilsson Hamiltonian is diagonalized in a set of states which become exact eigenstates only at large deformations, called the asymptotic basis wave functions \cite{Rassey,Quentin,Boisson}, which are characterized by the quantum numbers $K[\mathcal{N}n_z \Lambda]$, where $\mathcal{N}$ is the number of oscillator quanta, $n_z$ is the number of quanta along the cylindrical symmetry axis, $\Lambda$ and $K$ are the projections of the orbital angular momentum and the total angular momentum respectively along the same axis. These quantum numbers remain rather good even at intermediate deformation values \cite{Nilsson2}. Ben Mottelson \cite{Mottelson} has remarked that the asymptotic quantum numbers of the Nilsson model can be seen as a generalization of Elliott's SU(3), applicable to heavy deformed nuclei. 

A nontrivial question arising above is the following one: SU(3) is known to be the symmetry of the three-dimensional isotropic harmonic oscillator \cite{Wybourne,MS}. How is one allowed, to use it in the case of the deformed (anisotropic) harmonic oscillator with cylindrical symmetry \cite{Takahashi}? This is a formidable mathematical problem, which has already been addressed in the literature in several different ways, based on the fact that the $z$-component of the angular momentum operators remains intact by the transition from spherical to cylindrical symmetry \cite{RD,ND,PVI,Lenis,Sugawara,Arima}.
In particular, using quantum group techniques \cite{Chari,Klimyk,Daskal} it has been shown how the irreducible representations of the anisotropic HO with rational ratios of frequencies can be constructed, leaving the $z$-component of the angular momentum operator unchanged \cite{Lenis}. Furthermore, Smirnov and collaborators have shown how the isotropic HO can be transformed into the anisotropic oscillator through the use of a dilatation operator \cite{Asherova}. The resurrection of the SU(3) symmetry of the isotropic HO for large anisotropic deformations has been proved numerically in \cite{SugawaraArima}. 

We have described above some basic nuclear shapes (spherical, $\gamma$-unstable, axially deformed) and the symmetries corresponding to them, which are U(5), O(6), and SU(3) respectively. There are also nuclei exhibiting behavior intermediate among these symmetries \cite{AI3}. Considering various series of isotopes, one can see a gradual development from one symmetry into another. If at some point this transition becomes abrupt, we say that we have a shape/phase transition \cite{IZ,IJMPB,BonMcC100,RMP82}. Such an abrupt change is seen in the transition from vibrational (U(5)) to $\gamma$-unstable (O(6)) nuclei, called the E(5) critical point symmetry \cite{IacE5,CZE5,BonE5,RoweNPA,Turner}. Another case appears in the transition from spherical (U(5)) to axially deformed prolate (SU(3)) nuclei, called the X(5) shape phase transition \cite{IacX5,CZX5,BonX5,Rosensteel}. According to the Ehrenfest classification \cite{IJMPB}, E(5) is a second order phase transition, while X(5) is a first order one, i.e., it is more abrupt. As a consequence, it is much easier to locate experimental examples for X(5) than for E(5) (see the review articles \cite{McCutchan,BonRRP,CastenPPNP} for specific experimental manifestations of these critical point symmetries). Extensive literature exists \cite{RMP82} on shape phase transitions, both within the Bohr collective framework and the Interacting Boson Model approach. 

A peculiar situation, which might be related to the concept of shape phase transitions \cite{Heyde69,Ramos}, occurs in several even-even nuclei, in which the ground state band is accompanied 
by another $K=0$ band, which lies close in energy but possesses a radically different structure 
\cite{WoodPR,HeydeRMP}. This effect, which also appears in odd nuclei \cite{HeydePR}, is called shape coexistence and is attracting recently wide interest. In particular, shape coexistence is known to appear in certain areas of the nuclear chart,  while it seems to be absent in others (see Fig. 8 of \cite{HeydeRMP}). The arise of the additional coexisting $K=0$ band is usually attributed to two-particle--two-hole (and, more generally, n-particle--$n$-hole) excitations across a nuclear closed shell \cite{WoodPR,HeydeRMP}. Within the Bohr Hamiltonian approach, the two $K=0$ coexisting bands are considered to live within two different minima of a sextic potential \cite{Budaca,Georgoudis}.
{\it However, no explanation has been provided yet for the borders of the regions, within which shape coexistence is observed.}

It is the main purpose of the present chapter to show a path towards an explanation of shape coexistence in terms of a mechanism based on the two different sets of magic numbers (shell model, 3D-HO) mentioned above, taking advantage in parallel of the exact SU(3) symmetry present in the 3D-HO shells and the approximate proxy-SU(3) symmetry present in the shell model. In addition to predicting (free of any free parameters) the borders of the regions of coexistence, this method provides also specific predictions for regions of coexistence of a prolate with an oblate band, or of a prolate band with another prolate band of significantly different deformation. The collective model deformation parameters characterizing each band are also predicted by the theory, without involving any free parameters.

\section{The origin of QQ interaction}\label{1}
 A nucleus can exhibit collective features only if there is a kind of interaction between distant nucleons. So there must be a long range effective potential to unite the whole shell, or even different shells. The most important term of this long range potential is proportional to the quadrupole interaction.

Suppose there is a kind of two-body central force. The Taylor expansion of it shall be \cite{Harvey}:
\begin{equation}
V^{(2)}=\sum_{i<i'}V\left({r_{ii'}\over \alpha}\right)=\sum_{i<i'}\left(\xi _0+\xi _2{r_{ii'}^2\over \alpha^2}+\xi _4{r_{ii'}^4\over \alpha^4}+...\right),
\end{equation}
where $\alpha$ is a range parameter. 

The first term is dominant for a very long range potential, which for an A-body problem is approximately $\xi _0[A(A-1)/2]$. The second term simply verifies that the average potential of all nucleons, can be represented by an isotropic harmonic oscillator potential. The third term is \cite{Harvey}:
\begin{equation}
\sum_{i<i'}r_{ii'}^4=\sum_{i<i'}\left[r_i^4+r_{i'}^4+{8\over 3}r_i^2r_{i'}^2-4(r_i^2+r_{i'}^2)r_ir_{i'}\cos{\theta _{ii'}}+{4\over 3}r_i^2r_{i'}^2P_2(\cos{\theta _{ii'}})\right].
\end{equation}
The first three terms in the above expansion do not contribute any splitting in the energy degeneracy problem. Therefore the only significant term is the ${4\over 3}r_i^2r_{i'}^2P_2(\cos{\theta _{ii'}})$, where $P_2$ is the Legendre polynomial. This important term can be written as \cite{Harvey}:
\begin{equation}
r_i^2r_{i'}^2P_2(\cos{\theta _{ii'}})=(r_i^2Y_{20}(\theta _i\phi _i))\cdot (r_{i'}^2Y_{20}(\theta _{i'}\phi _{i'})),
\end{equation}
where $Y_{20}$ is the spherical harmonic with $l=2, m=0$. 

But the spherical harmonics are related to the collective quadrupole moment. The collective operators have the following relation with the space variables \cite{Harvey}:
\begin{equation}\label{Yq}
r_i^2Y_{2m}(\theta _i\phi _i)={b^2\over 4}\sqrt{5\over \pi}q^{(c)}_m,
\end{equation}
where  b is the oscillator length parameter with $b=\sqrt{\hbar \over m \omega}$. The difference between q and Q is that the first stands for one nucleon, while the capital for the whole nucleus.

So it happens that \cite{Harvey}:
\begin{equation}
r_i^2r_{i'}^2P_2(\cos{\theta _{ii'}})=\left({b^2\over 4}\sqrt{5\over \pi}\right)^2q^{(c)}_0q^{(c)}_0.
\end{equation}
This result proves, that any central potential contains a long range effective QQ interaction.

\section{The Nilsson Model}

Nilsson model is a microscopic nuclear model, in the spirit of shell model, applicable in deformed nuclei. The following will be a review of eq. (2)-(7) of \cite{Nilsson1}. The Hamiltonian for the nuclear coordinate system is:
\begin{eqnarray}
H=H_0+u_{ls}\hbar\omega_0(\delta)\Lambda\cdot \Sigma+u_{ll}\hbar\omega_0(\delta)\Lambda^2,\\
H_0=-{\hbar ^2\over 2m}{\nabla '}^2+{m\over 2}(\omega _x^2{x'}^2+\omega_y^2{y'}^2+\omega_z^2{z'}^2).
\end{eqnarray}
The frequencies are set to be:
\begin{eqnarray}
\omega_x^2=\omega_y^2=\omega_0^2(\delta)\left(1+{2\over 3}\delta\right),\\
\omega_z^2=\omega_0^2(\delta)\left(1-{4\over 3}\delta\right).
\end{eqnarray}
The volume conservation restriction leads to the relation:
\begin{equation}\label{omega}
\omega_0(\delta)=\tilde \omega_0\left(1-{4\over 3}\delta^2-{16\over 27}\delta^3\right)^{-1/6}.
\end{equation}
The deformation parameter $\delta$ is connected with the deformation $\beta$ of Bohr and Mottelson model as:
\begin{equation}
\delta\approx {3\over 2}\sqrt{5\over 4\pi}\beta\approx 0.95 \beta.
\end{equation}

The dimensionless coordinate system can be created by setting:
\begin{equation}
x=\sqrt{m\tilde \omega_0\over \hbar}x',\qquad y=\sqrt{m\tilde \omega_0\over \hbar}y',\qquad z=\sqrt{m\tilde \omega_0\over \hbar}z'.
\end{equation}
Then the spatial part of the Hamiltonian becomes:
\begin{eqnarray}
H_0=\tilde H_0+H_\delta,\\
\tilde H_0=\hbar \omega_0(\delta){1\over 2}(-\nabla ^2+r^2),\\
H_\delta=-\delta \hbar \omega_0(\delta){4\over 3}\sqrt{\pi\over 5}r^2Y_{20}.
\end{eqnarray}
Therefore the spatial part of the Hamiltonian can be interpreted as an isotropic 3D harmonic oscillator with frequency $\omega_0(\delta)$ plus a distortion to the isotropic field $H_\delta$.

This distortion contains the term $r^2Y_{20}$, which (recall eq. (\ref{Yq})) is the $q^{(c)}_0$ component of the quadrupole moment of the nucleon. So:
\begin{equation}
H_\delta=-\delta \hbar \omega_0(\delta){b^2\over 3}q^{(c)}_0.
\end{equation}
This term is the quadrupole interaction of the nucleon with the mean quadrupole field of all the others. It is noticeable that in the Nilsson model only the $q_0$ component appears, while the other four ($q_{\pm 2}, q_{\pm 1}$) are neglected.

The non vanishing matrix elements of $H_\delta$ are among Nilsson states with \cite{Nilsson1}:
\begin{equation}
\Lambda=\Lambda', \qquad \Sigma=\Sigma', \qquad l=l',l'\pm2,\qquad \mathcal{N}=\mathcal{N}',\mathcal{N}'\pm 2,
\end{equation}
where $\Lambda, \Sigma$ are the projections of the angular momentum and spin respectively, $l$ is the angular momentum and N is the total number of quanta. Obviously the $q^{(c)}$ interplays with orbitals with the same parity, since they have $\mathcal{N}, \mathcal{N}\pm 2$.

The full Nilsson model Hamiltonian can therefore be written as:
\begin{equation}\label{NilssonH}
H=\tilde H_0+u_{ls}\hbar\omega_0(\delta)\Lambda\cdot \Sigma+u_{ll}\hbar\omega_0(\delta)\Lambda^2-\delta \hbar \omega_0(\delta){4\over 3}\sqrt{\pi \over 5}r^2Y_{20}.
\end{equation}

\subsection{The Nilsson basis}

The asymptotic wave functions of Nilsson orbitals are labeled usually as $K[\mathcal{N}n_z\Lambda]$ \cite{Nilsson2}. These wave-functions occur when the deformation $\epsilon$ is large enough. An elegant method for the derivation of the asymptotic wave-functions is presented in \cite{Nilsson2}. In the following emphasis will be given in an equivalent notation $\ket{n_zrs\Sigma}$ of the asymptotic wave-functions, which highlights the number of quanta in the $x-y$ plane \cite{Nilsson2}. 

The Nilsson problem is solved in the cylindrical coordinate system. The quanta on the x-y plane are being created by the $R^\dagger,R,S^\dagger,S$ operators \cite{MotNil} are exactly the same with the $a_+^\dagger,a_+, a_-^\dagger,a_-$ of section \ref{Elliott}:
\begin{equation}\label{RS}
R^\dagger=a_+^\dagger, \qquad R=a_+,\qquad S^\dagger=a_-^\dagger,\qquad S=a_-.
\end{equation}
It is valid that:
\begin{equation}
n_\perp=r+s=n_x+n_y,
\end{equation}
where $R^\dagger R\ket{r}=r\ket{r}$ and $S^\dagger S\ket{s}=s\ket{s}$. The actions of the creation and annihilation operators on the relevant states are \cite{Shankar}:
\begin{equation}\label{actions}
R\ket{r}=\sqrt{r}\ket{r-1}, \quad R^\dagger \ket{r}=\sqrt{r+1}\ket{r+1},\quad S\ket{s}=\sqrt{s}\ket{s-1},\quad S^\dagger \ket{s}=\sqrt{s+1}\ket{s+1}.
\end{equation}
In addition the absolute value of the projection of orbital angular momentum in Nilsson model is \cite{Nilsson2}:
\begin{equation}
\Lambda=r-s.
\end{equation}
Last, for the projection of total angular momentum K and the projection of spin $\Sigma$ the equation \cite{Nilsson2}:
\begin{equation}\label{K}
K=\Lambda+\Sigma,
\end{equation}
applies.

With these tools:
\begin{eqnarray}
r+s=\mathcal{N}-n_z,\\
r-s=\Lambda,\\
\Sigma=K-\Lambda,
\end{eqnarray}
one can transform the Nilsson orbitals between the two notations:
\begin{equation}
K[\mathcal{N}n_z\Lambda]\rightarrow \ket{n_zrs\Sigma}.
\end{equation}
Some transformations are given in Table \ref{notation}.

\begin{table}[htb]
\caption{Nilsson orbitals in two different notations}\label{notation}
\begin{center}
\begin{tabular}{c c c}
Shell model & $K[\mathcal{N}n_z\Lambda]$ & $\ket{n_zrs\Sigma}$\\
$1g_{9/2}$ & 9/2[404] &$ \ket{040+}$\\
& 7/2[413] & $\ket{130+}$ \\
& 5/2[422]& $\ket{220+}$\\
& 3/2[431] & $\ket{310+}$\\
& 1/2[440] &$ \ket{400+}$\\
$2p_{1/2}$ & 1/2[301]& $\ket{021-}$\\
$1f_{5/2}$& 5/2[303] & $\ket{030-}$\\
&3/2[301] &$ \ket{021+}$\\
 & 1/2[310]&$ \ket{111+}$\\
$2p_{3/2}$& 3/2[312]& $\ket{120-}$\\
& 1/2[321]& $\ket{210-}$\\
$1f_{7/2}$& 7/2[303]&$\ket{030+}$\\
& 5/2[312] & $\ket{120+}$\\
&3/2[321]& $\ket{210+}$\\
&1/2[330]& $\ket{300+}$\\
$1d_{3/2}$ & 3/2[202] & $\ket{020-}$\\
 & 1/2[200] & $\ket{011+}$\\
$2s_{1/2}$ & 1/2[211] & $\ket{110-}$\\
$1d_{5/2}$ & 5/2[202] & $\ket{020+}$\\
 & 3/2[211] & $\ket{110+}$\\
& 1/2[220] & $\ket{200+}$\\
$1p_{1/2}$ & 1/2[101] & $\ket{010-}$\\
$1p_{3/2}$ & 3/2[101] & $\ket{010+}$\\
 & 1/2[110] & $\ket{100+}$\\
$1s_{1/2}$ & 1/2[000]& $\ket{000+}$\\
\end{tabular}
\end{center}
\end{table}

\section{The Elliott SU(3)}\label{Elliott}
  The majority of physicists nowadays are familiar with the nuclear shell model, but few have been engaged in the Elliott SU(3) symmetry, which was first introduced in \cite{Elliott1, Elliott2, Elliott3}. {\it Briefly this symmetry is the fulfillment of the shell model. While the shell model treats single particle states, the Elliott SU(3) teaches us, how to couple the valence nucleons towards the derivation of the collective nuclear states.} Moreover the Elliott SU(3) has a remarkable and unique property: it is a fermionic, collective, nuclear model. As a consequence the collective states emerge in accordance with the Pauli principle. 

The nucleus consists of Z protons and N neutrons. The mass number is $A=Z+N$. All these nucleons create a mean field potential, which participates in the 3D isotropic, harmonic oscillator Hamiltonian: 
\begin{equation}\label{osc}
H=\sum_{i=1}^AH_i=\sum_{i=1}^A\left({p_i^2\over 2m}+{1\over 2}m\omega_i^2r_i^2\right).
\end{equation} 
The single particle eigenstates of:
\begin{equation}\label{Hi}
H_i={p_i^2\over 2m}+{1\over 2}m\omega_i^2r_i^2,
\end{equation}
are labeled by the total number of oscillator quanta $\mathcal{N}_i$ in the three spatial dimensions.

The single particle, isotropic, harmonic oscillator Hamiltonian can be transformed in the second quantization with the use of the quanta annihilation and creation operators in each Cartesian axis $a_j, a^\dagger_j,$ with $ j=x,y,z $. The Hamiltonian of the $i^{th}$ particle in the $x$ direction is $H_{x,i}={p_{x,i}^2\over 2m}+{1\over 2}m\omega_i ^2x_i^2$. The eigenstates are $\ket{n_x}$. The action of the $a_x,a^\dagger_x$ on the eigenstates is \cite{Shankar}:
\begin{equation}
a_x\ket{n_x}=\sqrt{n_x}\ket{n_x-1}, \qquad a^\dagger_x\ket{n_x}=\sqrt{n_x+1}\ket{n_x+1}.
\end{equation}
By changing the script $x$ to $y$ or $z$ one gets the relevant actions in the other two Cartesian directions. In addition the number operators are:
\begin{equation}
a_j^\dagger   a_j \ket{n_j}=n_j\ket{n_j}.
\end{equation}
The above operators being boson (quanta) operators follow the well known commutation relations:
\begin{equation}
[a^\dagger_j,a^\dagger_{j'}]=[a_j,a_{j'}]=0,\qquad [a_j,a^\dagger_{j'}]=\delta_{jj'},\quad \mbox{ for }, j,j'=x,y,z.
\end{equation}
Thus the single particle 3D Hamiltonian can be transformed as \cite{Shankar}:
\begin{equation}
H_i={p_i^2\over 2m}+{1\over 2}m\omega^2r_i^2=\left(a_xa_x^\dagger+a_ya_y^\dagger+a_za_z^\dagger+{3\over 2}\right)\hbar\omega_i,
\end{equation}
with eigenvalues:
\begin{equation}
E_i=\left(n_x+n_y+n_z+{3\over 2}\right)\hbar \omega_i=\left(\mathcal{N}_i+{3\over 2}\right)\hbar \omega_i.
\end{equation}

Now the problem can be turned in cylindrical coordinates, which are more suitable for axially deformed nuclei. In order to distinguish the x-y plane from the z axis the left and right quanta operators are being defined \cite{Cohen}:
\begin{eqnarray}
a_{\pm}={a_x\mp i a_y\over \sqrt{2}},\\
a_{\pm}^\dagger={a_x^\dagger\pm i a_y^\dagger\over \sqrt{2}}.
\end{eqnarray}
The operators $a_+,a_+^\dagger$ are called right hand operators, because they seem to destroy/create a circular quantum, which rotates like a right hand in the x-y plane \cite{Cohen}. For the same reason the other two operators are called left hand operators. 
The operators in the z axis are renamed:
\begin{equation}
a_0=a_z, \qquad a_0^\dagger=a_z^\dagger.
\end{equation}
Now the Hamiltonian is \cite{Lipkin}:
\begin{equation}
H_i=\hbar \omega_i\left(a_+^\dagger a_+ +a_-^\dagger a_-+a_0^\dagger a_0+{3\over 2}\right).
\end{equation}

With these ingredients one can define eight operators which are the generators of SU(3) \cite{Lipkin}:
\begin{eqnarray}
l_0=a_+^\dagger a_+-a_-^\dagger a_-,\label{g1}\\
l_\pm=\mp (a_0^\dagger a_\mp-a_\mp^\dagger a_0),\label{g2}\\
q_{\pm 2}=-\sqrt{6}a_\pm^\dagger a_\mp,\label{g3}\\
q_{\pm 1}=\mp\sqrt{3}(a_0^\dagger a_\mp+a_\pm ^\dagger a_0),\label{g4}\\
q_0=2a_0^\dagger a_0-a_+^\dagger a_+-a_-^\dagger a_-.\label{g5}
\end{eqnarray}
The first three operators ($l_0, l_\pm$) form the SU(2) algebra of angular momentum, while the last five are the five componets of the single particle quadrupole operator. 

The eight of them close the commutation relations of the SU(3) algebra \cite{Lipkin}.
\begin{eqnarray}
{[l_0,l_\pm]=\pm l_\pm},\\
{[l_+,l_-]=2l_0},\\
{[l_0,q_m]=mq_m},\\
{[l_\pm,q_m]=\sqrt{6-m(m\pm 1)}q_{m\pm 1}},\\
{[q_0,q_{\pm 1}]=\pm 3\sqrt{3}l_\pm},\\
{[q_1,q_{-1}]= -3l_0},\\
{[q_2,q_{-2}]=6l_0},\\
{[q_{\pm 2}, q_{\mp 1}]=\pm 3\sqrt{2}l_{\pm}},\\
{[q_0,q_{\pm 2}]=[q_{\pm 1}, q_{\pm 2}]}=0.
\end{eqnarray}
The commutator of two operators from the set (\ref{g1})-(\ref{g5}) results to one operator from the set (\ref{g1})-(\ref{g5}) again. Thus the 8 operators of (\ref{g1})-(\ref{g5}) close the SU(3) algebra.

In simple words the Elliott SU(3) consists of the three angular momentum operators and of the five components of the quadrupole operator. The eight generators of SU(3) are certain linear combinations of the boson operators $a_j^\dagger,a_j$. The 3D isotropic harmonic oscillator Hamiltonian is invariant under U(3) and SU(3) transformations. The $(\mathcal{N}+1)(\mathcal{N}+2)\over 2$ orbitals of a harmonic oscillator shell with $\mathcal{N}$ total number of quanta have SU(3) symmetry.

\subsection{Derivation of the highest weight irrep}\label{hw}

In the Elliott SU(3) model the calculation of every observable is based on the Elliott quantum numbers $(\lambda, \mu)$ \cite{Elliott1}. The highest weight (h.w.) irrep describes the ground state properties of the nucleus. The method for the calculation of the h.w. $(\lambda, \mu)$ has been presented in steps at \cite{UN, Rila}.

For the highest weight irreps a simple method will be presented here. \\
1. The Cartesian states $\ket{n_z,n_x,n_y}$ are vectors of a $U\left({(\mathcal{N}+1)(\mathcal{N}+2)\over 2}\right)$ algebra. The degenerate 3D isotropic harmonic oscillator orbitals are ordered as:
\begin{equation}
\ket{n_z,n_x,n_y}=\ket{n,0,0},\quad \ket{n-1,1,0},\quad \ket{n-1,0,1},\quad ...,\quad \ket{0,0,n}.
\end{equation}
For total quanta $\mathcal{N}=2$ there are ${(\mathcal{N}+1)(\mathcal{N}+2)\over 2}=6$ orbitals. One has to order these 6 orbitals as follows:
\begin{eqnarray}\label{kets}
\ket{1}=\ket{2,0,0}, \quad \ket{2}=\ket{1,1,0},\quad \ket{3}=\ket{1,0,1},\quad \ket{4}=\ket{0,2,0},\quad \ket{5}=\ket{0,1,1},\quad \ket{6}=\ket{0,0,2}.
\end{eqnarray}
2. Then at most 2 protons or neutrons are placed in each orbital $\ket{n_z,n_x,n_y}$ with the order of Eq. (\ref{kets}).\\
For instance $\ce{_{12}^{31}Mg_{19}}$ has $19-8=11$ neutrons in the 8-20 shell. Two neutrons are placed in each of the $\ket{1},...,\ket{5}$ orbitals, while the last neutron occupies the $\ket{6}$.\\
3. Afterwards the summations of the quanta in each Cartesian axis are being calculated:
\begin{equation}
\sum_{i=1}^{Z_{val}, N_{val}}n_{zi}, \qquad \sum_{i=1}^{Z_{val}, N_{val}}n_{xi},\qquad \sum_{i=1}^{Z_{val}, N_{val}}n_{yi}.
\end{equation}
For the 11 neutrons in $U(6)$ the above expressions give $8, 8, 6$ respectively.\\
4. The results of the summations are placed in decreasing order and these are the quantum numbers $[f_1,f_2,f_3]$ of the U(3) albegra. For our example $[f_1,f_2,f_3]=[8,8,6]$.\\
7. The $\lambda$ and $\mu$ are:
\begin{equation}
\lambda=f_1-f_2,\qquad \mu=f_2-f_3,
\end{equation}
which leads to $(\lambda, \mu)=(0,2)$ for 11 particles in U(6).

\subsection{The collective operators}

The single particle operators can be extended to the many particle ones, by summing over all valence particles \cite{Lipkin}:
\begin{eqnarray}
L=\sum_i^{A_{val}}l_i,\\
Q_m=\sum_i^{A_{val}}q_{im}.
\end{eqnarray}
The former one is the angular momentum, while the latter is the $m$ componet of the algebraic quadrupole moment for the whole shell. These many particle operators satisfy the same commutation relations with their relevant single ones and therefore they close an SU(3) algebra.

The simplest SU(3) Hamiltonian of the many nucleon problem, which is suitable for a collective $0^+$ state, is \cite{Smirnov}:
\begin{equation}\label{H}
H=\sum_i^AH_i-{1\over 2}\chi Q\cdot Q=H_0-{1\over 2}\chi Q\cdot Q,
\end{equation}
where $\chi$ is the strength of the algebraic quadrupole-quadrupole (QQ) interaction and if $i$ and $i'$ are two distinct valence nucleons:
\begin{equation}\label{QQ}
QQ=\sum_{i,i'}^{A_{val}}\sum_{m=-2}^2(-1)^mq_{mi}q_{-mi'}=\sum_{m=-2}^2(-1)^mQ_mQ_{-m}.
\end{equation}

The SU(3) algebra has a second and a third order Casimir operator \cite{Bible}:
\begin{eqnarray}
C_2=\lambda^2+\mu^2+\lambda\mu+3(\lambda+\mu),\\
C_3={2\over 9}(\lambda^3-\mu^3)+{1\over 3}\lambda\mu(\lambda-\mu)+(\lambda+2)(2\lambda+\mu).
\end{eqnarray}
The Casimir operators commute with all the generators of the algebra. The QQ interaction is measured via the $C_2$ as \cite{Bible}:
\begin{equation}
QQ=4C_2-3L^2.
\end{equation}
The 3D isotropic harmonic oscillator Hamiltonian $H_0$ has dimensionless eigenvalues (i.e., $\hbar \omega_i=1$):
\begin{equation}
N_0=\sum_{i=1}^A\left(\mathcal{N}_i+{3\over 2}\right)
\end{equation}
The strength $\chi$ is measured by the equation \cite{chi}:
\begin{equation}
{\chi \over 2}={\hbar \omega \over 4N_0},
\end{equation}
where $\hbar\omega$ is measured in MeV.

The deformation variables $\beta, \gamma$ follow the formulas \cite{beta}:
\begin{eqnarray}
\beta^2={4\pi\over 5(A\bar r^2)^2}(C_2+3),\\
\gamma=\tan^{-1}{\sqrt{3}(\mu+1)\over 2\lambda+\mu+3},
\end{eqnarray}
where $A$ is the mass number and the dimensionless mean square radius is $\bar r^2=0.87^2A^{1/3}$ 
\cite{Stone}. The $\beta^2$ may be multiplied by the scaling factor $(A/S)^2$, where $S$ is the size of the proton and neutron valence shell \cite{proxy2}:
\begin{equation}
\beta^2={4\pi\over 5(S\bar r^2)^2}(C_2+3).\\
\end{equation}

\subsection{The $\bf SU(3)\rightarrow SU(2)\times U(1)$ decomposition}\label{decomp}

\begin{figure}[ht]
\begin{center}
\caption{Cylindrical symmetry. The $SU(3)$ algebra is decomposed into an $SU(2)$, which describes the 
$x$-$y$ plane and into a $U(1)$, which refers to the $z$ axis. This decomposition suits perfect to a prolate or oblate nuclear shape.}\label{cyl}
\includegraphics[scale=0.05]{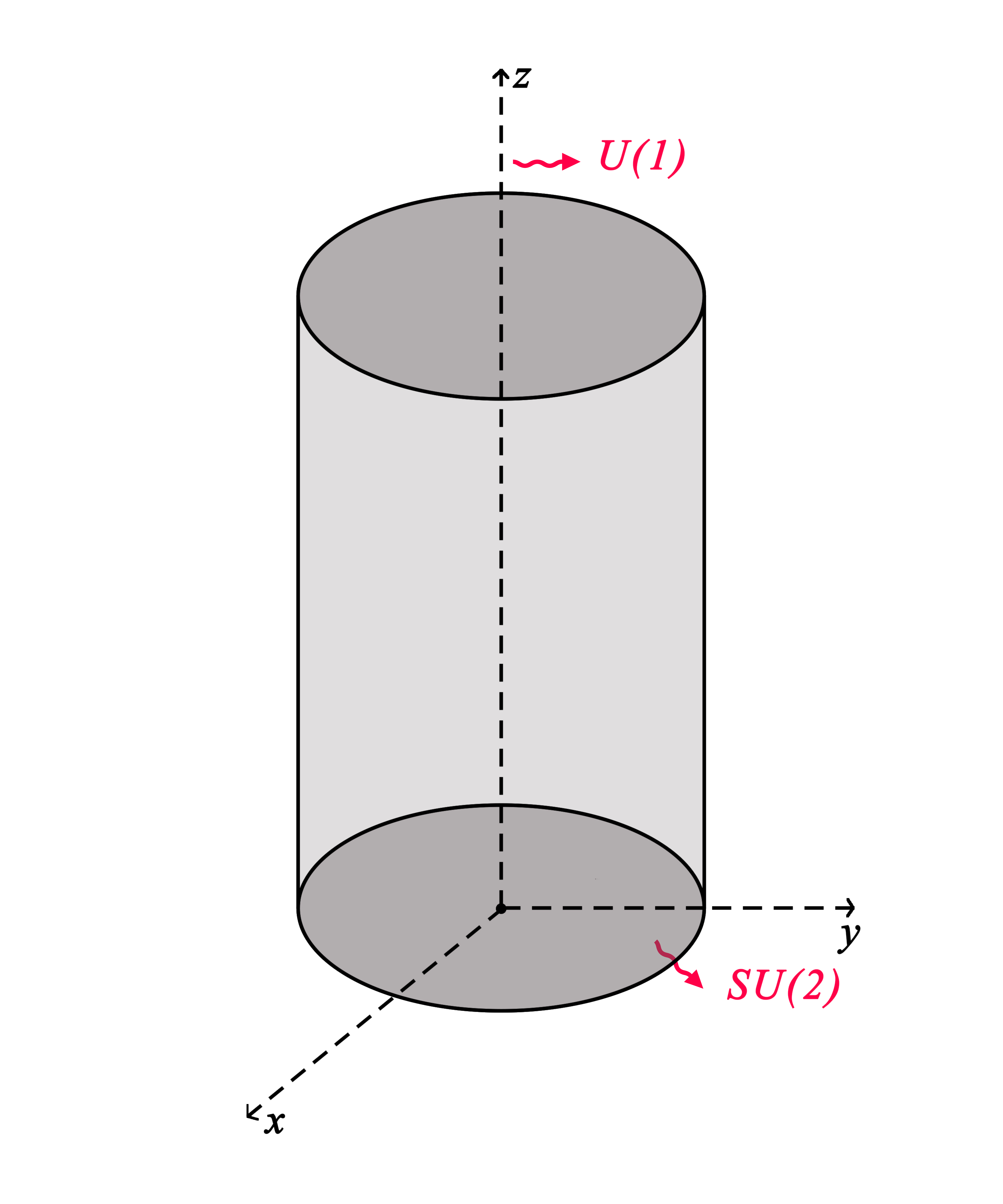}
\end{center}
\end{figure}

Most nuclei have quadrupole deformation. Such nuclei look like an ellipsoid with cylindrical symmetry. The cylinder is divided into the $x$-$y$ plane and the symmetry $z$ axis. If the ellipsoid has elongation across one axis ($z$), the shape is prolate, while if the elongation is across two axes ($x,y$), it is oblate. 

The algebra SU(3) is decomposed into two subalgebras SU(2) and U(1) \cite{Elliott2}. The former represents rotations in the $x$-$y$ plane, while the latter involves the $z$ axis. The U(1) is characterized by the operator $q_0$, which has eigenvalues $q_0=2n_z-n_\perp,$ with $n=n_z+n_\perp$. 

The SU(2) algebra has the following generators \cite{Elliott2}: 
\begin{equation}
u_{\pm}=\mp \left({1\over 2\sqrt{3}}\right)q_{\pm 2}, \qquad u_0={1\over 2}l_0,
\end{equation}
which satisfy the commutation relations:
\begin{equation}
[u_{\pm},u_0]=\mp u_{\pm},\qquad  [u_+,u_-]=-u_0.
\end{equation}
The SU(2) multiplets are characterized by a quantum number of $j$-type and one of 
$m_j$-type \cite{Bible}. In the present algebraic structure the $m_j$-type quantum number is the eigenvalue of $u_0$ and the $j$-type, let it be named $\Lambda_E$ ($\Lambda$ of Elliott), is
\begin{equation}\label{LE}
 \Lambda_E={1\over 2}(n_x+n_y)={1\over 2}n_\perp.
\end{equation}
Since:
\begin{equation}\label{u0}
 u_0={l_0\over 2},
\end{equation}
and the possible values of $u_0$ are $-\Lambda_E$, $-\Lambda_E +1$, \dots,
$\Lambda_E$ obeying the angular momentum algebra, obviously: 
\begin{equation}
l_0=-2\Lambda_E, -2\Lambda_E +2, \dots,2\Lambda_E.
\end{equation}
The basis of the $x-y$ plane is labeled as $\ket{\Lambda_E,u_0}$. These vectors with the use of (\ref{LE}) and (\ref{u0}) are more conveniently named $\ket{n_\perp,l_0}$. \\

{\it As a consequence the SU(2) subalgebra of the Elliott SU(3) depends solely on the quanta on the $x-y$ plane. The physical quantity of the SU(2) is the projection of the orbital angular momentum ($l_0$), while the $q_{\pm 2}$ are useful as ladder operators. The number of quanta in the z axis are isolated in the U(1) subalgebra (see Fig. (\ref{cyl})). This decomposition of the Elliott SU(3) suits perfectly in nuclei with quadrupole deformation.}\\

\begin{table}[htb]
\caption{Nilsson orbitals for the nucleons between 20 and 50. Due to the $ls$ interaction the $1f_{7/2}$ orbital is excluded from the shell under discussion, while the $1g_{9/2}$ orbital is included. Thus the harmonic oscillator shell 20-40, becomes the nuclear shell 28-50. The 28-50 shell consists of some orbitals with $\mathcal{N}=3$ and some orbitals with $\mathcal{N}=4$ quanta \cite{Casten}. Therefore the Elliott SU(3) symmetry is not valid in the 28-50 shell.}\label{ls}
\begin{center}
\begin{tabular}{c c c}
Shell model & $K[\mathcal{N}n_z\Lambda]$ & $\ket{n_zrs\Sigma}$\\
\color{blue}$1g_{9/2}$ & \color{blue}9/2[404] & \color{blue}$\ket{040+}$\\
& \color{blue}7/2[413] & \color{blue}$\ket{130+}$ \\
&\color{blue} 5/2[422]&\color{blue} $\ket{220+}$\\
& \color{blue}3/2[431] & \color{blue}$\ket{310+}$\\
& \color{blue}1/2[440] &$ \color{blue}\ket{400+}$\\
$2p_{1/2}$ & 1/2[301]& $\ket{021-}$\\
$1f_{5/2}$& 5/2[303] & $\ket{030-}$\\
&3/2[301] &$ \ket{021+}$\\
 & 1/2[310]&$ \ket{111+}$\\
$2p_{3/2}$& 3/2[312]& $\ket{120-}$\\
& 1/2[321]& $\ket{210-}$\\
\color{violet}$1f_{7/2}$& \color{violet}7/2[303]&\color{violet}$\ket{030+}$\\
& \color{violet}5/2[312] &\color{violet} $\ket{120+}$\\
&\color{violet}3/2[321]&\color{violet} $\ket{210+}$\\
&\color{violet}1/2[330]&\color{violet} $\ket{300+}$\\
\end{tabular}
\end{center}
\end{table}

\begin{table}[htb]
\caption{The $1f_{7/2}$ orbital has been excluded from the harmonic oscillator shell 20-40 due to the $ls$ interaction, while the $1g_{9/2}$ orbital has intruded in it. The $1f_{7/2}$ and $1g_{9/2}$ orbitals consist of Nilsson orbitals with $\mathcal{N}=3$ and $\mathcal{N}=4$ respectively, as seen in Table \ref{ls}. These orbitals are listed here using the $\ket{n_zrs\Sigma}$ notation. All of them have common projection of spin $\Sigma$. Each of the orbitals of $1f_{7/2}$ has identical distribution of quanta in the $x$-$y$ plane $(\ket{rs})$ with an orbital of $1g_{9/2}$. Such orbitals appear in the same line in the Table and are depicted in the same color. The orbitals with the same color differ by one quantum in the $z$ axis. Therefore the operators $q_{\pm 2}, l_0$ of the SU(2) subalgebra of SU(3) (section \ref{decomp}) give exactly the same eigenvalues when acting on the colored orbitals. As a consequence the SU(2) operators cannot discriminate among the two sets of the colored orbitals \cite{hnps}. }\label{orbitals}
\begin{center}
\begin{tabular}{c c c c}
Shell Model & $\ket{n_zrs\Sigma}$ & Shell Model & $\ket{n_zrs\Sigma}$\\
$1g_{9/2}$ & $\ket{040+}$ & & X \\ 
& $\ket{1{\color{blue}30+}}$ & $1f_{7/2}$ & $\ket{0{\color{blue}30+}}$  \\
& $\ket{2{\color{violet}20+}}$ & & $\ket{1{\color{violet}20+}}$ \\
& $\ket{3{\color{red}10+}}$ & & $\ket{2{\color{red}10+}}$\\
& $ \ket{4{\color{green}00+}}$ & & $\ket{3{\color{green}00+}}$\\
\end{tabular}
\end{center}
\end{table}

\begin{table}[htb]
\caption{The replacement of the orbitals in proxy-SU(3). One quantum from the $z$ axis is erased from each intruder orbital, while the $n_x$, $n_y$ (or, equivalently, the $r$, $s$) of the intruders and the normal parity orbitals remain intact. The intruder with the highest projection of total angular momentum ($K$ in Nilsson notation) is excluded. The new shell (after the approximation) is a complete harmonic oscillator shell, thus it has an SU(3) symmetry.}\label{Proxy}
\begin{center}
\begin{tabular}{c c c c c}
 & Before & & After &\\
shell model & $\color{blue}K[\mathcal{N}n_z\Lambda]$ & $\color{blue}\ket{n_zrs\Sigma}$& $\color{violet}K[\mathcal{N}n_z\Lambda]$ & $\color{violet}\ket{n_zrs\Sigma}$ \\
$1g_{9/2}$ &\color{blue} 9/2[404]& $\color{blue}\ket{040+}$ & \color{violet}X & \color{violet}X\\
&\color{blue} 7/2[413] & $\color{blue}\ket{130+}$ &\color{violet}7/2[303]  & $\color{violet}\ket{030+}$  \\
& \color{blue}5/2[422]& $\color{blue}\ket{220+}$  & \color{violet}5/2[312]  & $\color{violet}\ket{120+}$\\
& \color{blue}3/2[431] & $\color{blue}\ket{310+}$ &\color{violet}3/2[321]  & $\color{violet}\ket{210+}$ \\
& \color{blue}1/2[440] & $\color{blue}\ket{400+}$  & \color{violet}1/2[330]  & $\color{violet}\ket{300+}$\\
$2p_{1/2}$ & 1/2[301] & $\ket{021-}$ & 1/2[301] & $\ket{021-}$\\
$1f_{5/2}$& 5/2[303] & $\ket{030-}$ & 5/2[303] & $\ket{030-}$ \\
&3/2[301] & $\ket{021-}$ &3/2[301] & $\ket{021-}$  \\
 & 1/2[310] & $\ket{111+}$& 1/2[310] & $\ket{111+}$\\
$2p_{3/2}$& 3/2[312] & $\ket{120-}$ & 3/2[312] & $\ket{120-}$\\
& 1/2[321] & $\ket{210-}$ & 1/2[321] & $\ket{210-}$ \\
\end{tabular}
\end{center}
\end{table}

\section{Proxy-SU(3) symmetry}\label{sym}

A nucleus consists of $Z$ protons and $N$ neutrons. The nucleons interact mainly via the strong nuclear force. Since the exact formula for this force is not yet known, one can approach the result of all the inter-nucleon interactions as in section \ref{1}. The main result of all the inter-nucleon interactions is the creation of a mean field potential, which is described by a 3D isotropic harmonic oscillator potential. This mean field creates nuclear shells, which consist of orbitals with common number of oscillator quanta $\mathcal{N}$. The magic numbers of these shells are the 3D harmonic oscillator (HO) magic numbers: 2, 8, 20, 40, 70, 112, ...

Unfortunately the mean field is not the only interaction. There exists a single particle spin-orbit (SO)  term $ls$, which destroys the harmonic oscillator shells. The $ls$ interaction forces the orbitals to move in energy. As an example, the 20-40 harmonic oscillator shell becomes 28-50. The mechanism of this procedure is the following:\\
a) The harmonic oscillator shell 20-40 consists of the shell model orbitals $1f_{7/2}, 2p_{3/2}, 1f_{5/2}, 2p_{1/2}$. The orbital $1g_{9/2}$ is part of the harmonic oscillator shell 40-70. The orbitals are listed in Table \ref{ls}.\\
b) The $ls$ interaction lowers the energy of the $1f_{7/2}$ orbital so much, that this orbital is no longer included in the shell under discussion.\\
c) Instead the $1g_{9/2}$ from the 40-70 HO shell is lowered and becomes part of the shell under discussion.\\
To resume, the harmonic oscillator shell 20-40 loses the $1f_{7/2}$ orbital and gains instead the $1g_{9/2}$ orbital. This new shell lies between magic numbers 28 and 50.\\

{\it Thus  after the $ls$ interaction is applied, each harmonic oscillator shell with $\mathcal{N}$ quanta loses the orbital with total angular momentum $j=\mathcal{N}+{1\over 2}$ and gains the orbital with $\mathcal{N}+1$ quanta and $j'=\mathcal{N}+1+{1\over 2}$.}\\

Proxy-SU(3) was introduced in \cite{proxy1} in order to describe a nuclear shell which consists of some orbitals with $\mathcal{N}$ quanta and some orbitals with $\mathcal{N}+1$ quanta. Nuclear shells above 28 protons or neutrons are of this type. Since such shells do not match with an exact SU(3) symmetry, an approximation is necessary in order to make the algebraic tools applicable. The proxy-SU(3) replacement of orbitals is justified in Table \ref{orbitals} and described in Table \ref{Proxy} for the 28-50 shell.\\

{\it The proxy-SU(3) approximation affects only the orbitals with $\mathcal{N}+1$ quanta (intruder orbitals), by erasing one quantum from the $z$ axis of each intruder orbital. The normal parity orbitals remain intact. }

\subsection{The exact symmetry behind proxy-SU(3)}

Proxy-SU(3) has been very successful in predicting for various isotopic chains, without any parameters, the $\beta$ and $\gamma$ deformations of each nucleus, as well as the prolate over oblate dominance along with the prolate-oblate transition \cite{proxy2}. This success relies on the highest weight irrep and on the fact that there is an exact symmetry behind proxy-SU(3).

The total wave function of a single particle state is the product of a spatial and a spinor part. The spatial wave function is characterized by the number of quanta in the $z$ axis ($\ket{n_z}$) and in the $x$-$y$ plane ($\ket{rs}$). In a full harmonic oscillator shell with $\mathcal{N}$ quanta the $n_z$ gets values:
\begin{equation}
n_z=\mathcal{N},\mathcal{N}-1,\mathcal{N}-2,...,0.
\end{equation}
The quanta in the $x$-$y$ plane are respectively:
\begin{equation}
r+s=0,1,2,...,\mathcal{N},
\end{equation}
with the restriction \cite{Nilsson2}:
\begin{equation}
r\ge s.
\end{equation}
Thus the following combinations are valid:
\begin{eqnarray*}
(n_z=\mathcal{N}, \quad r=s=0),\\
(n_z=\mathcal{N}-1, \quad r=1, \quad s=0),\\
(n_z=\mathcal{N}-2, \quad r=2, \quad s=0),\\
(n_z=\mathcal{N}-2, \quad r=1, \quad s=1),\\
...
\end{eqnarray*}
As an example the orbitals of the ``Before'' column of Table \ref{Proxy}, excluding the 9/2[404] orbital, contain all the $r$, $s$ values of a full harmonic oscillator shell with $\mathcal{N}=3$ quanta.\\

{\it Consequently a nuclear shell, such as 28-50, 50-82, 82-126, after excluding the Nilsson orbital with the maximum value of $K$, has the SU(2) subalgebra of the Elliott SU(3). The proxy-SU(3) approximation is not affecting the number of quanta in the $x$-$y$ plane. Thus the orbitals of the proxy-SU(3) shell preserve the exact SU(2) symmetry.}

\begin{table}[htb]
\caption{A significant spin-orbit interaction may create the SO-like shells 2-6, 6-14. Proxy-SU(3) can be applied in these shells. The column ``Before'' refers to the original nuclear shell and the column ``After'' to the proxy-SU(3) shell \cite{hnps}.}\label{Proxy2}
\begin{center}
\begin{tabular}{c c c c}
 & Before & After & \\ 
Shell Model & $K[\mathcal{N}n_z\Lambda]$ & $K[\mathcal{N}n_z\Lambda]$ & SO-like magic number\\
$1f_{7/2}$& 7/2[303]& X  & 28\\
& 5/2[312] & 5/2[202]& \\
&3/2[321]& 3/2[211]&  \\
&1/2[330]& 1/2[220]&\\
$1d_{3/2}$& 3/2[202]& 3/2[202] & \\
& 1/2[200]& 1/2[200] & \\
$2s_{1/2}$&  1/2[211]& 1/2[211] &\\
&&&\\
$1d_{5/2}$& 5/2[202]& X &14\\
& 3/2[211]& 3/2[101]& \\
& 1/2[220]& 1/2[110]& \\
$1p_{1/2}$& 1/2[101]& 1/2[101] &  \\
&&&\\
$1p_{3/2}$& 3/2[101]& X & 6\\
& 1/2[110]& 1/2[000] &  \\
&&&\\
$1s_{1/2}$& 1/2[000]&1/2[000] & 2\\
\end{tabular}
\end{center}
\end{table}

\section{Magic numbers below 28}\label{mn}
       In section \ref{sym} it has been described how the spin-orbit interaction creates the nuclear shell 28-50. By the same procedure the shells 50-82, 82-126 are created. But a significant spin-orbit term in the Hamiltonian can create magic numbers below 28 particles, as described in Table \ref{Proxy2}, which we are going to call, following Ref. \cite{Sorlin}, SO-like magic numbers. The SO-like magic numbers, that will arise, are:
\begin{equation}
\mbox{{\it SO-like magic numbers below 28}: }2, 6, 14.
\end{equation}
Actually the SO-like magic number 14 is already well established experimentally \cite{Sorlin}.

{\it Clearly the above SO-like magic numbers have not emerged in this work through the study of energy gaps, but through symmetry considerations. The sets of orbitals among the mentioned SO-like magic numbers, after excluding the Nilsson orbital with maximum $K$, preserve the SU(2) symmetry of Elliott. }

As seen in Table \ref{Proxy2} the SO-like shells 6-14, 14-28 have a proxy-SU(3) symmetry. These shells can be treated as in \cite{proxy1, proxy2}, to calculate the deformation variables $(\beta, \gamma)$. The shell 14-28 is useful for the neutrons of the $\ce{Mg}$ isotopes, while the 6-14 can be applied in the neutrons of the $\ce{Be}$ isotopes, to predict the parity inversion at $\ce{^{11}Be}$.

\begin{table}[htb]
\caption{The highest weight SU(3) irreps \cite{proxy2} for  13 to 28 nucleons, listed in the case in which the SO-like magic numbers 14 and 28 and the proxy-SU(3) symmetry are taken into account, as well as in the case in which the harmonic oscillator magic numbers 8, 20, and 40 and the HO SU(3) symmetry are considered.}\label{irreps}
\begin{center}
\begin{tabular}{c c c c c c}
Even & & & Odd & &\\
Particle Number & $(\lambda, \mu)_{SO}$&$(\lambda, \mu)_{ho}$& Particle Number & $(\lambda, \mu)_{SO}$&$(\lambda, \mu)_{ho}$\\
14 & (0, 0) & (6, 0) & 13& (0, 0) & (5, 1)\\
16 & (4, 0) & (2, 4) & 15 & (2, 0) & (4, 2)\\
18 & (4, 2) & (0, 4) & 17 & (4, 1) & (1, 4)\\
20 & (6, 0) & (0, 0) & 19 & (5, 1) & (0, 2)\\
22 & (2, 4) & (6, 0) & 21 & (4, 2) & (3, 0)\\
24 & (0, 4) & (8, 2) & 23 & (1, 4) & (7, 1)\\
26 & (0, 0) & (12, 0) & 25 & (0, 2) & (10, 1)\\
28 & (0, 0) & (10, 4) & 27 & (0, 0) & (11, 2)\\
\end{tabular}
\end{center}
\end{table}

\begin{figure}[htb]
\flushleft{
\caption{Shape coexistence begins when the deformation obtained with the harmonic oscillator magic numbers is less than the deformation calculated with the SO-like magic numbers, {\it i.e.} $\beta_{ho}\le\beta_{SO}$ and usually stops at a harmonic oscillator magic number. The mechanism of \cite{PhD, hnps, Rila} predicts that the nuclei $\ce{_{12}^{30-32}Mg_{18-20}}$ are candidates for shape coexistence. 
Data for $\beta$ have been taken from \cite{Tables}. The data indicate that 
 the neutrons of the $\ce{Mg}$ isotopes with neutron number $N\le 17$ follow the harmonic oscillator magic numbers and do not exhibit shape coexistence. As a consequence, when $N\le 17$, there is a unique ground state obeying the rules of the 8-20 HO shell. But as soon as shape coexistence begins, $N>17$, there may be two states low lying in energy. As shown in Eq. (\ref{st}), in general the state with the higher deformation is the ground state. As seen from the comparison to the data \cite{Nudat2}, the ground state is derived from the SO-like shell 14-28. The excited state is derived in general from the coupling of the 8-20 HO shell with the 14-28 SO-like shell. To summarize, when $N>17$ the ground state follows the 14-28 SO-like shell, while when $N<17$ the ground state follows the 8-20 HO shell. This may be the cause of the inversion of states, which begins at $\ce{_{12}^{30}Mg_{18}}$ \cite{Scheit}. }\label{betaMg}
\includegraphics[scale=0.4]{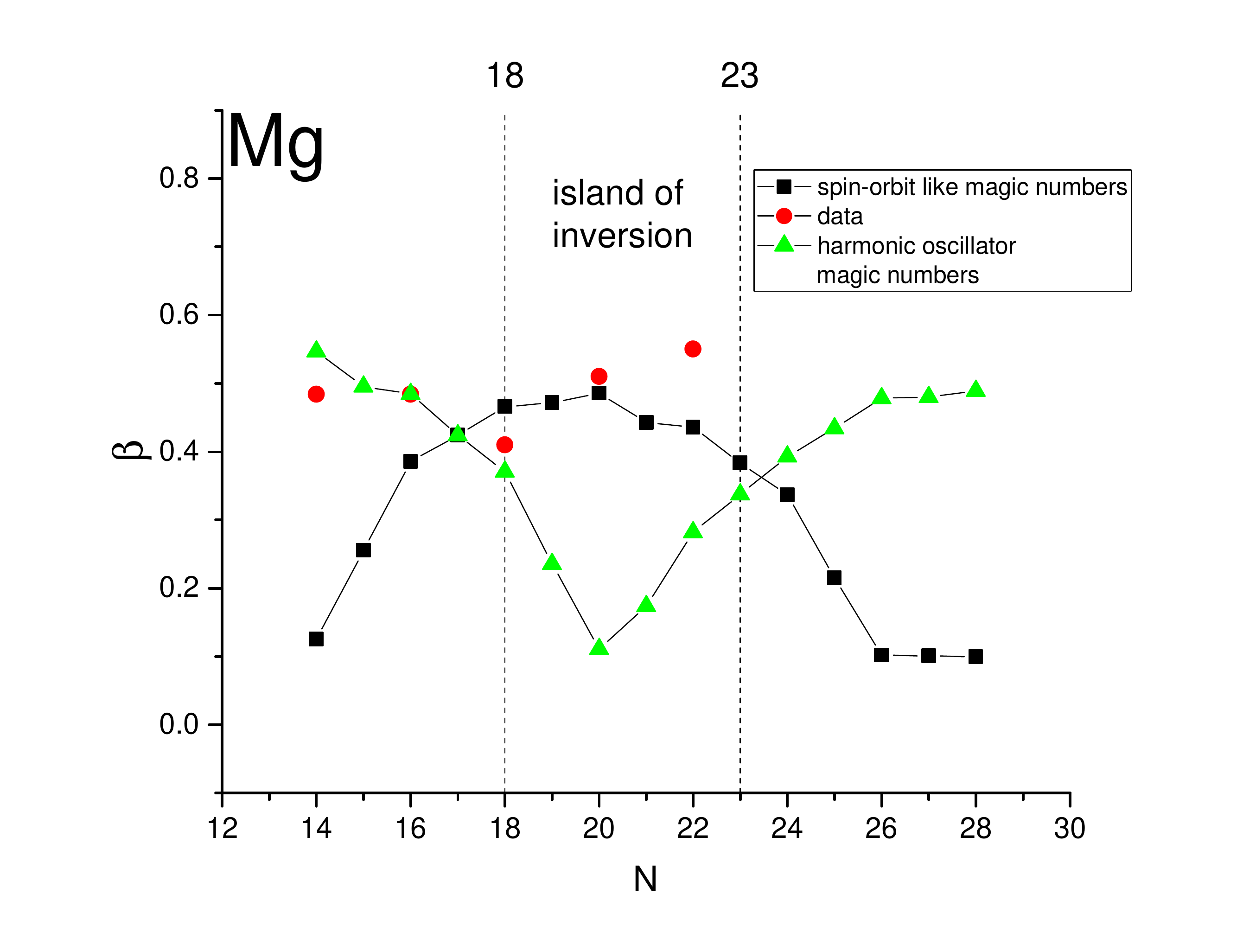}}
\end{figure}

\section{Magic numbers $\rightarrow$ Shape coexistence $\rightarrow$ Inversion of states}

The $\ce{Mg}$ isotopic chain is the best example, which demonstrates how the SO-like magic numbers below 28 may lead to right predictions about deformation $\beta$, shape coexistence and inversion of states. A bulk of experimental work \cite{Scheit, 32Mg, 30Mg, moments1, moments2, moments3} has established that the $\ce{Mg}$ isotopic chain exhibits both shape coexistence and inversion of states. A shell model mechanism, which uses particle-hole excitations, is the state of the art theoretical approach for these phenomena in $\ce{Mg}$ \cite{Merge, rich, Poves}. 

Detailed calculations will be presented for $\ce{_{12}^{26-40}Mg_{14-28}}$ within the proxy-SU(3) framework. In the following a mechanism, which was initially proposed in \cite{PhD}, will be used to explain why:\\
a) $N=20$ is not a magic number for the ground state of $\ce{_{12}^{32}Mg_{20}}$,\\
b) shape coexistence occurs among neutron numbers $N=18-20$ for this isotopic chain, and\\
c) inversion of states appears at $\ce{_{12}^{31}Mg_{19}}$.

The isotonic chains with $N\approx 20$, possess values of the deformation $\beta$ which exhibit a minimum  at $Z=14$ \cite{Nudat2}. This provides experimental support for the findings of the previous section, in which 14 was found to be a SO-like magic number, for protons in the present case. The 12 protons of $\ce{Mg}$ lie in the 6-14 shell, and in proxy-SU(3) have $(\lambda, \mu)_p=(0,0)$. 

The neutrons  of $\ce{_{12}^{26-40}Mg_{14-28}}$ lie either in the SO-like shell 14-28, or in the harmonic oscillator shells 8-20 and 20-40. The highest weight SU(3) irreps for these shells are listed in Table \ref{irreps}. Since protons do not contribute at all, the deformation formula becomes:
\begin{equation}
\beta^2={4\pi\over 5(S_n\bar r^2)^2}(C_2+3),\\
\end{equation}
where $\bar r^2=0.87^2N^{1/3}$  \cite{Stone}, $C_2=\lambda_n^2+\mu_n^2+\lambda_n\mu_n+3(\lambda_n+\mu_n)$ (where the subscript $n$ stands for neutrons) and $S_n=12$ for the 8-20 and the 14-28 proxy-SU(3) shell, while $S_n=20$ for the 20-40 shell. The theoretical predictions for $\beta$ for the $\ce{Mg}$ isotopic chain for the spin-orbit like magic numbers and the harmonic oscillator magic numbers are presented in Figure \ref{betaMg}.

Shape coexistence is a phenomenon where the nucleus exhibits a ground state band with a certain type of deformation (prolate for instance) and a slightly excited band with another type of deformation (oblate for example) \cite{HeydeRMP, HeydePR, WoodPR}. The idea, that has been proposed in \cite{PhD, hnps}, is that shape coexistence, is the consequence of two sets of magic numbers, namely the SO-like 6, 14, 28, 50, 82, 126 magic numbers and the harmonic oscillator 2, 8, 20, 40, 70, 112 magic numbers: \\

{\it When the nuclear deformation with the harmonic oscillator magic numbers becomes less than the deformation with the SO-like magic numbers: 
\begin{equation}\label{coex}
\beta_{ho}\le\beta_{SO},
\end{equation}
then shape coexistence is likely to appear.}\\

 The ground state of such nuclei derives from the SO-like magic numbers, because this shell possesses the maximum deformation and $QQ$ interaction, as we can see using the Hamiltonian of Eq. (\ref{H}). Indeed, schematically one has  
\begin{eqnarray}
H_{ho}=H_0-{\chi\over 2}QQ_{ho}, \\
H_{SO}=H_0-{\chi\over 2}QQ_{SO},\\
QQ_{ho}\le QQ_{SO}\Rightarrow H_{SO}\le H_{ho}.\label{st}
\end{eqnarray}
The excited $K=0^+$ state of nuclei with shape coexistence derives from the coupling of the harmonic oscillator shell 8-20 with the SO-like shell 14-28. In the review articles \cite{HeydeRMP,WoodPR} one can see that the experimentally known examples of shape coexistence in several series of isotopes appear below and stop at a harmonic oscillator magic number. For example, in the Hg series of isotopes, shape coexistence appears below and stops at $N=112$, as seen in Fig. 10 of Ref. \cite{HeydeRMP}). In the present case this means that shape coexistence should appear below and stop at $\ce{_{12}^{32}Mg_{20}}$.

The inversion of states is a side effect of shape coexistence. One can divide the $\ce{Mg}$ isotopic chain in two halves:\\
a) the first half contains the isotopes with neutron numbers $N\le17$ with $\beta_{ho}\ge\beta_{SO}$, and\\
b) the second contains the isotopes with $17<N<23$ with $\beta_{ho}<\beta_{SO}$.\\
The neutrons of the isotopes of the first half clearly follow the harmonic oscillator magic numbers 8-20 (see Fig. \ref{betaMg}). But the isotopes of the second half may exhibit shape coexistence. Such isotopes have two low lying states: a ground state and a slightly excited one. \\

{\it While the ground state of $\ce{Mg}$ isotopes with $N\le 17$ followed the 8-20 HO shell, now the ground state of isotopes with $N>17$ follows the rules of the 14-28 SO-like shell. The inversion of states may be caused by a change of magic numbers.} \\

The cause of  the inversion of states may be hidden in the collective nuclear features instead of the single particle ones.     The Elliott SU(3) model contains all the necessary techniques, which can reveal the $J^\pi$ of the ground state and of the excited state of even-odd nuclei, which lie within the islands of inversion. 

\section{Conclusions}

A mechanism is proposed for shape coexistence and inversion of states \cite{PhD, hnps, Rila}. The mechanism involves two sets of magic numbers: the spin-orbit (SO) like magic numbers 6, 14, 28, 50, 82, 126 and the harmonic oscillator magic numbers 2, 8 20, 40, 70, 112. Shape coexistence arises when $\beta_{SO}\ge \beta_{ho}$ and stops at a harmonic oscillator closure. The state with the maximum deformation lies lower in energy. As a consequence, within the islands of shape coexistence shown in Fig. 8 of Ref. \cite{HeydeRMP} the ground state is derived by the SO-like magic numbers, while the coexisting excited 
$0^+$ state is derived from the coupling of the harmonic oscillator with the SO-like magic numbers. The inversion of states is a side effect of shape coexistence. The proposed mechanism predicts without any parameters the borders of the islands of shape coexistence appearing at Fig. 8 of \cite{HeydeRMP}. This mechanism can be applied in every mass region to predict the borders of the islands of shape coexistence and all the relevant nuclear observables.

\end{document}